\documentstyle[11pt]{article}

\begin{document}


\begin{center}
{\Large\bf A glimpse at the flat-spacetime limit of quantum gravity\\
using the Bekenstein argument in reverse\footnote{This essay received
an ``honorable mention" in the 2004 Essay Competition of the Gravity
Research Foundation -- Ed.}}

\end{center}

\vskip 0.5 cm
\begin{center}
{\bf Giovanni AMELINO-CAMELIA}$^a$, {\bf Michele ARZANO}$^b$,
{\bf Andrea PROCACCINI}$^a$\\
\end{center}

\begin{center}
{\small $^a${\it Dip.~Fisica Univ.~Roma ``La Sapienza''
and Sez.~Roma1 INFN, }
{\it Piazzale Moro 2, Roma, Italy}}\\
{\small $^b$ {\it Dept Physics and Astronomy,
 University of North Carolina,
 Chapel Hill, NC 27599, USA}}
\end{center}

\vskip 0.5 cm
\begin{center}
{\bf ABSTRACT}
\end{center}

\begin{quotation}
\leftskip=0.6in \rightskip=0.6in

{\small
\noindent
An insightful argument for a linear relation between the entropy and the
area of a black hole was given by Bekenstein using only
the energy-momentum dispersion relation,
the uncertainty principle, and
some properties of classical black holes.
Recent analyses within String Theory
and Loop Quantum Gravity describe black-hole entropy
in terms of a dominant contribution, which indeed depends linearly on the
area, and a leading log-area correction.
We argue that, by reversing the
Bekenstein argument, the log-area correction can provide insight
on the energy-momentum dispersion relation and the uncertainty principle
of a quantum-gravity theory. As examples we consider the energy-momentum
dispersion relations that recently emerged in the Loop Quantum Gravity
literature and the Generalized Uncertainty Principle that is expected
to hold in String Theory.
}
\end{quotation}

\baselineskip 17.5pt plus .5pt minus .5pt

As first observed by Bekenstein~\cite{bek},
the fact that
the entropy of a black hole should be proportional
to its (horizon-surface) area, up to
corrections that can be neglected when the area $A$ is much larger
than the square of the Planck length $L_p$,
can be derived using very simple ingredients.
In (classical) general-relativity one finds~\cite{christo}
that the minimum increase of area when the black hole absorbs a classical
particle of energy $E$ and size $s$ is $\Delta A \simeq 8 \pi L_p^2 E s$
(in ``natural units" with $\hbar=c=1$).
Taking into account the quantum properties of particles
one can estimate the size of the particle as roughly given by
its position uncertainty $\delta x$, and one can assume that
a particle with position uncertainty  $\delta x$
should at least~\cite{landau} have energy $E \sim  1/ \delta x$.
This leads to the conclusion\cite{bek,hod} that the minimum change
in the black-hole area must be of order $L_p^2$,
independently of the size of the area.
Then using the fact
that, also independently of the size of the area,
this minimum increase of area should correspond to the
minimum (``one bit") change of entropy, $(\Delta S)_{min} = \ln 2$,
one easily obtains~\cite{bek} the proportionality between
black-hole entropy and area.

Over the last few years
both in String Theory and in Loop Quantum Gravity
some techniques for the analysis of entropy on the basis of
quantum properties of black holes were developed.
These results\cite{stringbek,lqgbek} now go even beyond the
entropy-area-proportionality contribution:
they establish that the leading correction should
be of log-area type, so that one expects (for $A \gg L_p^2$)
an entropy-area relation for black holes of the type
\begin{equation}
S = \frac{A}{4 L_p^2}
+ \rho \ln \frac{A}{L_p^2} + O\left(\frac{L_p^2}{A}\right)
~.
\label{linPLUSlog}
\end{equation}
While all analyses agree on the coefficient of the term $A/L_p^2$,
different quantum-gravity theories appear to give
rise~\cite{stringbek,lqgbek,nonUniv}
to different predictions for the coefficient $\rho$ of the
leading $\ln (A/L_p^2)$ correction.

Here we propose that the log-area leading correction might admit
a simple description, just like the argument presented by
Bekenstein in Ref.~\cite{bek} gives a simple description
of the dominant linear-in-$A$ contribution.
Our simple description will also explain why
the coefficient $\rho$ of the log-area term takes different
values in different quantum-gravity theories.
And we stress that the availability of results on
the log-area correction
might provide motivation for reversing the
Bekenstein argument: the knowledge of the black-hole entropy-area
law up to the leading log correction can be used to establish the
Planck-scale modifications of the ingredients of the Bekenstein
analysis.

We start by observing that, as stressed above,
a key point for the Bekenstein analysis
is the  $E \ge 1/ \delta x$ relation between the energy of a particle
and the uncertainty in its position.
One obtains $E \ge 1/ \delta x$
by combining appropriately~\cite{landau}
the Heisenberg uncertainty principle,
the special-relativistic energy-momentum dispersion relation, and
the fact that a procedure for the measurement of the position of
a particle must not introduce an energy uncertainty for the particle
which is greater than its mass (in its rest frame), since
otherwise the procedure is subject to the possibility of production
of additional copies of the particle.
But nearly all approaches to the quantum-gravity problem predict
a Planck-length-dependent modification of $E \ge 1/ \delta x$,
since various quantum-gravity results suggest that
the uncertainty principle and/or the energy-momentum
dispersion relation should be modified.
Modifications of the uncertainty principle
have been discussed in quantum-gravity
theories that predict a new role for the Planck length in the localization
of particles, as in the case of String Theory~\cite{venegross}.
Several scenarios for a modified
energy-momentum dispersion relation~\cite{grbgac}
have been considered, particularly
in quantum pictures of spacetime which involve some level
of spacetime discreteness~\cite{thooftdiscrete}
or noncommutativity~\cite{majrue,kpoinap,dsr,aad03}.
In Loop Quantum Gravity, which indeed
predicts a certain form of spacetime granularity
(discrete spectrum of areas and
volumes), several studies\cite{lqgDispRel1,lqgDispRel2}
found evidence for a modification of the dispersion relation.

In order to accommodate the differences between alternative quantum-gravity
scenarios we consider a rather general parametrization of the
Planck-length modified $E \ge 1/ \delta x$ relation:
\begin{equation}
E \ge \frac{1}{\delta x} - \eta_1 \frac{L_p}{(\delta x)^2}
- \eta_2 \frac{L_p^2}{(\delta x)^3} +
O\left(\frac{L_p^3}{(\delta x)^4}\right)
\label{deform}
\end{equation}
We will argue that a log-area leading correction
to the entropy-area relation requires $\eta_1 = 0$. For $\eta_1 \neq 0$
the leading correction turns out to
behave like the square-root of the area.
A result in agreement with (\ref{linPLUSlog}) is obtained
for $\eta_1 = 0$, $\eta_2 \sim \rho$
(and the sign convention in (\ref{deform}) turns out to
ensure that $\eta_2$
and $\rho$ are either both positive or both negative).

As in the original Bekenstein argument~\cite{bek},
we take as starting point the general-relativity
result which establishes that the area
of a black hole changes according to $\Delta A \ge 8 \pi E s$
when a classical particle of energy $E$ and size $s$ is absorbed.
In order to describe the absorption of a quantum particle one must
describe the size of the particle in terms of the uncertainty
in its position~\cite{bek,hod}, $s \sim \delta x$, and take into account
a ``calibration
factor"~\cite{chenproc,calib2,calib3} $(\ln 2)/2 \pi$
that connects the $\Delta A \ge 8 \pi E s$ classical-particle result
with the quantum-particle
estimate $\Delta A \ge 4 (\ln 2) L_p^2 E \delta x$.
Following the  original Bekenstein argument~\cite{bek} one then
enforces the relation $E \ge 1/ \delta x$
(and this leads to $\Delta A \ge 4 (\ln 2) L_p^2$),
but we must take into account the Planck-length modification in (\ref{deform}),
obtaining
\begin{eqnarray}
\Delta A \ge 4 (\ln 2) \! \left[L_p^2  - \!  \frac{\eta_1 L_p^3}{\delta x}
- \! \frac{\eta_2 L_p^4}{(\delta x)^2} \right] \! \simeq
 4 (\ln 2) \! \left[ L_p^2  - \! \frac{\eta_1 L_p^3}{R_S}
- \! \frac{ \eta_2 L_p^4}{(R_S)^2}  \right]
\! \simeq  4 (\ln 2) \! \left[ L_p^2
- \! \frac{\eta_1 2 \sqrt{\pi} L_p^3}{\sqrt{A}}
- \! \frac{\eta_2 4 \pi L_p^4}{A}  \right]
\nonumber
\end{eqnarray}
where we also used the fact that in falling in the black hole
the particle acquires~\cite{calib2,smaller,bigger} position
uncertainty $\delta x \sim R_S$, where $R_S$ is the Schwarzschild
radius (and of course $A = 4 \pi R_S^2$).

Next, following again Bekenstein\cite{bek}, one assumes that the entropy
depends only on the area of the black hole, and one uses the fact
that according to information theory
the minimum increase of entropy should be $\ln 2$,
independently of the value of the area:
\begin{equation}
\frac{dS}{dA} \simeq \frac{min (\Delta S)}{min (\Delta A)}
\simeq  \frac{\ln 2}{4 (\ln 2)  L_p^2 \left[1
- \eta_1 2 \sqrt{\pi} \frac{L_p}{\sqrt{A}} -  \eta_2 4 \pi \frac{L_p^2}{A} \right]}
\simeq  \left(\frac{1}{4 L_p^2} +  \frac{\eta_1 \sqrt{\pi}}{2 L_p \sqrt{A}}
+  \frac{\eta_2  \pi}{A}\right)
~.
\label{minDa}
\end{equation}
From this one easily obtains (up to an irrelevant constant contribution
to entropy):
\begin{equation}
S \simeq \frac{A}{4 L_p^2}
+ \eta_1  \sqrt{\pi} \frac{\sqrt{A}}{L_p}
+ \eta_2  \pi \ln \frac{A}{L_p^2}
~.
\label{final}
\end{equation}
As anticipated the sought agreement with (\ref{linPLUSlog})
requires $\eta_1 = 0$ and $\eta_2 \sim \rho$ (precisely $\eta_2 =  \rho /\pi$).

From this we conjecture that
the log-area contribution to the black-hole entropy might be deeply
connected with a corresponding $L_p^2/(\delta x)^3$
correction to the $E \ge 1/ \delta x$ relation.
And, as anticipated, our description
encourages the possibility that the coefficient of the log-area correction
might be different in different quantum-gravity theories.
The dominant linear-in-$A$ term must be ``universal" since it
was obtained in Ref.~\cite{bek} using only observations
that do not involve any quantum-gravity effects (the Planck length appears in the
Bekenstein formula only as a result of the classical-gravity aspects
of the analysis, which of course involve the gravitational constant).
Instead the coefficient of the log-area leading correction
depends directly on the coefficient of the $L_p^2/(\delta x)^3$
correction to the $E \ge 1/ \delta x$ relation, and,
since different quantum-gravity
theories could lead to different modifications of the Heisenberg
uncertainty principle
and of the energy-momentum dispersion relation,
this coefficient should not be expected to take the same value
in all quantum-gravity theories.

Amusingly this observation provides an opportunity to use
the Bekenstein argument in reverse. In the 1970s
there was no result on entropy from the analysis
of quantum properties of black holes and analyses ``{\it a la} Bekenstein",
using some simple ingredients,
allowed to derive the linear term in the entropy-area formula.
Now that the analysis of quantum properties of black holes allows
to derive also the log-area correction we might be able to use this information
to infer which Planck-scale modifications
should be introduced in the ingredients of the Bekenstein analysis.

A first application of
this ``reversed Bekenstein argument" can be found
in Loop Quantum Gravity.
There is a large number of Loop-Quantum-Gravity studies
(see, {\it e.g.}, Refs.~\cite{lqgDispRel1,lqgDispRel2} and references therein)
reporting results in support of the possibility of a Planck-scale modified
energy-momentum dispersion relation, although
there is still no consensus on the precise form of this modification.
The ideal framework for establishing Planck-scale modifications of the dispersion
relation is the study of the self-energy of a particle in a low-energy effective
theory on ``quasi-Minkowski
spacetime" (a quantum spacetime which most closely
approximates Minkowski spacetime in a given quantum-gravity theory).
But in Loop Quantum Gravity the
well-known ``classical-limit problem"~\cite{lqgCLprob},
which of course includes a ``Minkowski-limit problem",
has provided so far a rather serious obstacle for
the analysis of Planck-scale modifications of dispersion relations.
In particular, two key alternatives
for the Planck-scale modification
of the energy-momentum dispersion relation are
still being considered~\cite{lqgDispRel1,lqgDispRel2}:
one in which the function $p(E)$ admits an expansion
with leading Planck-scale correction of order $L_p E^3$,
\begin{equation}
 \vec{p}^2 \simeq E^2 - m^2 + \alpha_1 L_p E^3
~,
\label{disprelONE}
\end{equation}
and one in which the function $p(E)$ admits an expansion
with leading Planck-scale correction of order $L_p^2 E^4$,
\begin{equation}
 \vec{p}^2 \simeq E^2 - m^2 + \alpha_2 L_p^2 E^4
~.
\label{disprelTWO}
\end{equation}
We intend to show that,
since also in Loop Quantum Gravity
there is evidence~\cite{lqgbek} that the leading correction to the entropy-area
black-hole formula is logarithmic,
our ``reversed Bekenstein argument"
favors the scenario (\ref{disprelTWO}).
The scenario (\ref{disprelONE}) would lead to a $\sqrt{A}$ leading correction.

On the basis of the analysis reported above it will be sufficient
to study the implications of these modified dispersion relations
for the relation between the energy of a particle and its position uncertainty,
and show that (\ref{disprelTWO}) leads to a contribution
of the type $L_p^2/(\delta x)^3$, while from (\ref{disprelONE})
one obtains a $L_p/(\delta x)^2$ term.

In order to establish the implications
for the relation between the energy of a particle and its position uncertainty,
we can follow the familiar derivation~\cite{landau} of the
relation $E \ge 1/ \delta x$, substituting, where applicable, the
standard special-relativistic dispersion relation with
the Planck-scale modified dispersion relation.
Let us first consider the case of
the dispersion relation (\ref{disprelTWO}).
It is convenient to start by focusing on the case
of a particle of mass $M$ at rest, whose position is being measured
by a procedure involving a collision with a photon of energy $E_\gamma$
and momentum $p_\gamma$.
In order to measure the particle position with precision $\delta x$
one should use a photon
with momentum uncertainty $\delta p_\gamma \ge 1/\delta x$.
This $\delta p_\gamma \ge 1/\delta x$ relation originates from Heisenberg's
uncertainty
principle for which, as mentioned, modifications have not been suggested
in the Loop-Quantum-Gravity  literature
(but they are expected in other approaches
to the quantum-gravity problem). Following the standard argument~\cite{landau},
one takes this $\delta p_\gamma \ge 1/\delta x$ relation and converts it into
the relation $\delta E_\gamma \ge 1/\delta x$, using the
special-relativistic dispersion relation,
and then the relation $\delta E_\gamma \ge 1/\delta x$ is converted into
the relation $M\ge 1/\delta x$ because the measurement procedure
requires\footnote{We are here again using the fact~\cite{landau}
that the measurement procedure must ensure
that the relevant energy uncertainties are not large enough
to possibly produce extra copies of the particle whose
position one intends to measure.} $M \ge \delta E_\gamma$.
If indeed Loop Quantum Gravity hosts a Planck-scale-modified
dispersion relation of the form (\ref{disprelTWO}),
it is easy to see that, following the same reasoning,
one would obtain from $\delta p_\gamma \ge 1/\delta x$
the requirement
\begin{equation}
M  \ge \frac{1}{\delta x} \left(1
 - \alpha_2 \frac{3 L_p^2}{2 (\delta x)^2}\right)
~.
\label{dis}
\end{equation}

These results strictly apply only to the measurement
of the position of a particle
at rest, but they can be straightforwardly
generalized~\cite{landau} (simply using a boost)
to the case of measurement
of the position of a particle of energy $E$.
In the case of the standard dispersion relation (without Planck-scale modification)
one obtains the familiar $E  \ge 1/\delta x$, as required for a purely linear
dependence of entropy on area.
In the case of (\ref{disprelTWO})
one instead easily finds
that $E  \ge \frac{1}{\delta x} \left(1- \alpha_2 \frac{3 L_p^2}{2
(\delta x)^2}\right)$,
which fulfills the requirements of our derivation of a leading-order correction
of log-area form.

The careful reader can easily adapt to the case of the
dispersion relation (\ref{disprelONE})
this simple analysis which we discussed
for the dispersion relation (\ref{disprelTWO}).
The end result is that from (\ref{disprelONE}) it follows
that $E  \ge \frac{1}{\delta x} \left( 1+ \alpha_1 \frac{L_p}{\delta x}\right)$,
and this in turn leads to a leading correction to
the entropy-area formula that
goes like the square-root of the area, rather than the predicted log-area
leading correction.
This confirms that, using the reversed Bekenstein argument,
the dispersion relation (\ref{disprelTWO}) should be preferred to
the dispersion relation (\ref{disprelONE}).

In closing let us turn to String Theory.
Also in String Theory there are robust indications of a logarithmic
contribution to the black-hole entropy-area relation,
but at present the mainstream String-Theory literature provides
no evidence for a Planck-scale modification of the dispersion relation.
In the case of String Theory our proposed ``reversed Bekenstein argument"
leads us to consider the so-called ``GUP"
or ``Generalized Uncertainty Principle",
\begin{equation}
\delta x  \ge \frac{1}{\delta p} + \lambda_s^2  \delta p
~,
\label{gup}
\end{equation}
which finds support in various String-Theory studies~\cite{venegross}.
The scale $\lambda_s$ in (\ref{gup}) is an effective string length
(a characteristic length scale of the GUP which should be closely
related to the string length and therefore the Planck length).

The presence of a logarithmic term in the entropy-area relation
can be viewed as a consequence of
the presence of the $\lambda_s^2  \delta p$ term in the
GUP. In fact, as the careful reader can easily verify,
from the GUP one obtains (following again straightforwardly
the standard measurability line of analysis~\cite{landau})
a modification of the relation $E  \ge 1/\delta x$.
The modification is of the type $E  \ge 1/\delta x + \Delta$,
with $\Delta$ of order $\lambda_s^2/\delta x^3$,
and originates from the fact that
according to the GUP, (\ref{gup}), one
obtains  $\delta p_\gamma \ge 1/\delta x +\lambda_s^2/\delta x^3$,
instead of the $\delta p_\gamma \ge 1/\delta x$
which follows from Heinsenberg's uncertainty principle.
As discussed above this type of relation
between the energy and the position uncertainty of a particle,
with a  $1/\delta x^3$
contribution in addition to the familiar $1/\delta x$ term,
leads, through the Bekenstein argument,
to a logarithmic contribution to the black-hole entropy-area
relation.

The examples of Loop Quantum Gravity and String Theory
show that we might have at our disposal
a rather powerful tool for the analysis of quantum-gravity theories,
assuming that it is legitimate to think of a connection between
the log-area contribution to black-hole entropy and
a $1/(\delta x)^3$
correction to the $E \ge 1/ \delta x$ relation.
Whereas in Loop Quantum Gravity
the $L_p^2/(\delta x)^3$ correction is motivated by results
on Planck-scale-modified dispersion relations
of the type (\ref{disprelTWO}),
in String Theory the analogous correction
of order $\lambda_s^2/\delta x^3$ is motivated by the GUP.
In general, in any given quantum-gravity theory,
the presence of a logarithmic term in the black-hole
entropy-area relation may be a manifestation of a
modified dispersion relation and/or a modified uncertainty principle.

\bigskip

\section*{Acknowledgments}
G.~A.-C.~gratefully acknowledges conversations
with O.~Dryer, D.~Oriti, C.~Rovelli and L.~Smolin.
The work of M.~A.~was supported by a Fellowship from The Graduate School of The
University of
North Carolina. M.~A.~also thanks the Department of Physics of the University of
Rome for hospitality.

\bigskip

\baselineskip 12pt plus .5pt minus .5pt

\end{document}